\documentclass[letterpaper]{article}
\usepackage{iccc}

\usepackage{dblfloatfix}
\usepackage{times}
\usepackage{helvet}
\usepackage{courier}
\usepackage{float}
\usepackage{array}
\usepackage{booktabs}
\usepackage{enumitem}
\usepackage{multirow}
\usepackage{amsmath}

\usepackage{tikz}
\usetikzlibrary{spy}
\usepackage{calc}

\newcommand{\comment}[1]{}
\newcommand{\removeComment}[1]{}

\title{Rock Climbing Route Generation and Grading as Computational Creativity}\author{Jesse Roberts\\
Computer Science Department\\
Vanderbilt University\\
Nashville, TN 37235\\
Jesse.Roberts@Vanderbilt.edu\\
}
\begin{document} 
\maketitle
\begin{abstract}
\begin{quote}

In this paper, we bridge work in rock climbing route generation and grading into the computational creativity community. We provide the necessary background to situate that literature and demonstrate the domain's intellectual merit in the computational creativity community. We provide a guiding set of desiderata for future work in this area. We propose an approach to computational route grading. Finally, we identify important gaps in the literature and consider how they may be filled. This paper thus also serves as a pilot study, planting a flag for our ongoing research in this domain. 

\end{quote}
\end{abstract}

\section{Introduction}

Rock climbing is a growing recreational sport, industry, and community. Climbing contributed 12 billion dollars to the national economy in 2017. In 2018, more than 4\% of Americans participated in rock climbing, and climbers contributed more than 65 thousand volunteer hours to conservation and management of climbing areas, most of which are public lands \cite{aac2019}. Climbing was included in the 2020 Tokyo Olympics for the first time and has been announced for continued inclusion in the 2024 Olympics with an increased number of medals allocated.

Most rock climbing today occurs indoors on routes that are created by route setters \cite{aac2019}. Setting these routes is a costly endeavor for climbing gyms as skilled setters must be well compensated for their expertise. For a gym to remain successful it must consistently set new routes. Here, there is an opportunity for computational systems to be developed that support this highly creative work. However, there are currently no available computationally creative or co-creative systems that can generate climbing routes in a goal-directed, human-consistent manner. 

This paper serves to bridge rock climbing (as a domain and body of literature) into the ICCC and to plant a flag for our reserach. We begin by examining how climbing routes are characterized and evaluated by the climbing community and introduce a computational interpretation. This is followed by a brief survey of the literature. From the review, we identify the need for domain-specific computationally creative systems that are capable of (1) inferring the most probable sequences of movements to be employed when ascending or traversing steep terrain, (2) considering the difficulty of the movement (grade) including psychological effects of fear, and (3) predicting how rewarding a sequence of movements may be. Computational critics of this variety can then be used to (4) perform goal-directed search for novel routes. These are summarized in Table \ref{tab:climbinggendesiderata}.

We survey the existing body of work toward computational climbing route generation and grading. We find that the literature has focused primarily on evaluating the climbing grade of a route (human consistent computational grading remains unsolved) in a highly reduced problem space. We believe the difficulty in grading may be due to important deficiencies in the literature regarding extraction and representation of basic, pertinent information (human consistent movement sequence) from the data (relative hold locations). However, computational grading may still fail to reach human consistency in general as the grade also captures psychological effects which have not been considered at all. Of the systems in the literature that attempt route generation, only one treats generated routes as creative artefacts to which an audience will have some reward response. None explicitly consider the style of the route.

Some climbing specific information may be stated without citation. One of the authors is a member of the rock climbing community, volunteers as a route setter as part of being the faculty sponsor for a collegiate rock climbing club, and is a member of the southern climbing coalition (a climber collective for purchasing, maintaining, and providing public access to climbable cliff lines in the south-eastern United States).

\comment{
However, of particular note is a fundamental gap in the path planning literature. Very little is known regarding how humans path plan in vertical terrain, with no substantive datasets available in the literature. Without an understanding of vertical path planning, it is impossible to generate routes that necessitate a desired beta (term for the sequence of movements employed by a climber). Without the ability to generate a route which necessitates a desired beta, any computationally generated route falls short of emulating the creative goals of a human route setter.
}

\section{Background}

Rock climbing as a discipline started in the mid 19\textsuperscript{th} century among mountaineers looking to hone a subset of skills required in technical ascents like those typical of the Italian Dolomites and the Swiss Alps \cite{zhu2021origin}. There were important areas which required would be ascenders to climb vertically using features in the rock as hand and foot holds. 

Outdoor rock climbing on natural holds is now a part of a larger sport that also includes indoor rock climbing on hard plastic ``holds" as seen in the 2020 Olympics. These holds can be bolted to a fabricated wall to set a climbable indoor ``route" which may be reset after a period of time (usually a few weeks). Rock climbing is no longer a sub-discipline of mountaineering but is now a distinct sport based in the mental and physical activity required to find and execute a path up a route.

Route setters create rock climbing routes using the wall as the canvas and the various shapes and locations of holds as the media. The setter designates the start hold(s), the holds that are available for use, and the hold on which the climber must end. Climbers then try to find a set of movements (path planning) that will allow them to establish, ascend, and finish the route. Solving an interesting ``problem" and executing an interesting sequence of movements provides the reward to the climber.

Over time, climbers grow in ability as they acquire climbing specific techniques and strength. So, routes of varying grade (term for difficulty) are needed for gyms to serve climbers of differing ability. Thus, setters must create routes that are not only rewarding to climbers, but that also correspond to desired grades. Commercial gyms employ teams of setters who constantly replace routes on a rotational schedule as reward begins to decay with repeated climbs on the same route (after successful ascension).

\subsection{The Creativity of Route Setting}

Just as the value of a painting is not in the distribution of color on a canvas but rather the response evoked in the observer, in rock climbing, what is valuable is not the route (even though the route is the tangible artefact), but is rather the experience that is facilitated/necessitated by the route. Climbers obtain reward from the experience by finding and executing a sequence of movements that allows them to ascend the route. However, some sequences of movement are more rewarding than others. For this reason, setters place holds on the wall to create a route that requires a sequence of movements and techniques for ascension/traversal to be (humanly) possible, much like a choreographer designing a dance. The art of setting, then, is in creating a route that requires climbers to execute a movement sequence which is rewarding.

Once set, a route takes on a life of its own through community interaction. The sequence of movements which the setter intended to be necessary (the author's intent) is not necessarily the sequence employed by climbers. This is because successfully climbing a route only requires that the climber start on the designated hold(s) and reach the top by using the holds provided by the setter. For this reason, route setting is all the more difficult. It is insufficient to create a route that \textit{permits} an interesting sequence of movements, as the path of least resistance will be the path employed by climbers. Therefore, the route must \textit{necessitate} a rewarding sequence of movements.

In practice this requires the setter to iteratively adjust the location of holds and climb the route to ensure the intended sequence is rewarding, can't be subverted, and matches the desired grade. This process is time consuming, physically difficult (due to the repeated climbs), and fundamentally artistic \cite{anderson_rennak_2004}.

\section{Characterizing Climbing Routes}

Climbing routes are primarily characterized by the difficulty or grade of the route. There are multiple conventions for the manner in which the grade is conveyed. However, in the United States, the Yosemite decimal system (YDS) is typical. 

The YDS splits terrain into classes numbered 1 to 5. Class 1 terrain refers to walk-able hiking, while class 5 refers to vertical free climbing. Rock climbing routes are always in the class 5 category. However, within this class, there are sub-classes designated by a decimal. The easiest rock climbing routes are rated 5.1 while the hardest are rated 5.15. There are further subdivisions for the grades from 5.10 to 5.15. These are designated by a letter following the grade with ascending lexicographic order corresponding to ascending difficulty ie. 5.12a is easier than 5.12b which is easier than 5.13a. 

While the YDS is standardized, the climbing grades (5.1 to 5.15d) are interesting as their assignment and use represents emergent behavior within the climbing community. They are not standardized or governed. Thus, their meaning and use have evolved.

\begin{figure}
    \centering
    \resizebox{!}{18em}{%
    \input{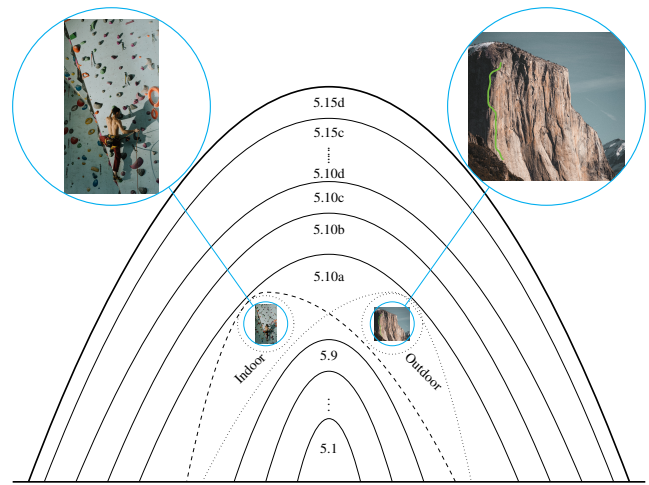}}
    \caption{Relation of climbing grades. Climbing a route from a given grade implies a high probability of success on other routes within the grade.}
    \label{fig:climbingGrade}
\end{figure}

The grade of a route is a ``heuristic" evaluation of the route's difficulty with no community accepted expectations for what features a route of a given grade does or does not include. That is, grades are not prescriptive. Rather, the climbing grade is best understood as a direct consequence of climbers engaging in computing with words to communicate relative probability of success. As shown in figure \ref{fig:climbingGrade}, the climbing grade conveys an expectation that a climber, capable of climbing a subset of routes of a certain grade, will be able to climb other routes of the same or lower grade. ie. if a climber is capable of climbing the first 2 pitches of the west buttress (5.10a shown on the right in Figure \ref{fig:climbingGrade}) on El Capitan, then the same climber has a high probability of climbing the unnamed, 5.10a indoor route shown on the left in Figure \ref{fig:climbingGrade} assuming that the climber's career exposure to possible route styles is a uniform random variable. 

While there are at times minor disagreements regarding a route's grade (which we hypothesize may largely stem from non-uniform climbing style exposure), the degree of uniformity of experienced difficulty of a climb across the community is very surprising. One would expect relative difficulty to be something individual rather than largely communal.

On the surface, this sounds simple and intuitive. However, all computational systems in the literature fail to achieve human consistent grade predictions. We believe the difficulty arises from the non-prescriptive nature of the grade. Rather than representing feature content of a route which may contribute difficulty (and thus being something that could be inferred based on route features), the grade of a route represents something altogether human. The grade captures the difficulty experienced by humans when they interact with the route, regardless of the cause of the difficulty.

\subsubsection{Fear \& difficulty}

Routes may induce more or less fear depending on the structure of the route. Routes that are overhung (have an angle which is greater than $90^{\circ}$) cause the climber to be more acutely aware of the height and exposure. Further, routes may require the climber to be in body configurations that are perceived as more or less dangerous (whether they are or not). The fear that is evoked during the climb contributes to the overall conditional probability of success as it is shown to affect the performance of the climber \cite{green2014impact}.

\subsubsection{Computational Interpretation}

From a set theoretic point of view, we can say that each grade defines a set of routes. A route, $R$, is defined to be a member of a grade set, $S_{5.\alpha}$, iff the probability of climbing the route given that the climber has climbed other routes within the set, ie. $P(R|S_{5.\alpha})$, is high AND the probability of climbing other routes within the set given success on the route, ie. $P(S_{5.\alpha}|R)$, is also high. So, 

$$R \in S_{5.\alpha} \iff P(R|S_{5.\alpha}) \land P(S_{5.\alpha}|R)$$

The above definition is intentionally vague regarding the meaning of "high" as this the nature of computing with words. 

Assigning a grade to a new route is precisely the problem of identifying the set to which the route belongs. However, there is no clear way to know the ground truth for all members of the community. Instead, climbers who have made early ascents of a route attempt to cluster it into the appropriate grade based on their experience. The specific clustering algorithm used is unknown and is likely different for each climber. This connotes all the more interest in the relative success of this system. 

\subsection{Computational Grading}

Potentially successful systems for predicting the climbing grade will need to cluster routes into grades in a manner that is informed by the meaning of the grade (discussed above). Essentially, the algorithm should take as input the established grades and routes and attempt to assign a grade to the route being graded such that a monoidal T-Norm \removeComment{\cite{esteva2001monoidal}} conjunction of $P(R|S_{5.\alpha})$ and $P(S_{5.\alpha}|R)$ is maximized.

The only work in the literature attempting to statistically analyze the appropriate grade for a given climb is \cite{scarff2020estimation}. Here the authors attempt to do this based on the whole history rating (WHR) of the climb, searching for a set of ratings for climbers and routes that maximizes the prediction of successful ascents. However, this is not equivalent to the above. Therefore, the clustering algorithm that is suggested here is novel. 

The dissimilarity comes from long standing routes not being considered at ground truth. A successful grading algorithm should allow for ordering effects otherwise it will fail to be consistent with reality. By not locking the grade once a route has received a certain level and length of community exposure, all grades are considered tentative which is not generally consistent with climbing community behavior.

\subsection{Style}

So far, it has been shown that the climbing grade (the most common characterization of climbing routes) is a non-prescriptive, relative conditional probability that captures all contributing aspects, including but not limited to the physical and emotional experience of the route. However, it is important to also discuss the non-difficulty based characteristics of the movement sequence required for the route. 

\comment{
\subsubsection{Visual Style}

Routes are artefacts of creative processes. As such, they possess authorial intent. It has been discussed that the ability to arrange holds on a wall without awareness if the climbing experience and beta is insufficient to meaningfully create routes. However, while a route setter is creating a climbing experience, they are also creating something that can be visually experienced. For this reason, route setters not only consider the climbing experience but may also consider the visual impact of the holds on the wall.

A computationally creative or co-creative system for the augmentation of route setting should ultimately have the ability to consider multiple objectives. As an example, given an ability to emulate human path planning and limitations a computational route setter should be able to add arbitrary holds to a given route without affecting the beta of the route by choosing holds and their placement that have no effect on the path. }

When creating a route, setters may choose to set the route such that the sequence of movements necessarily contains certain common moves, building the rest of the route around this inspiration. As an example, a route setter may start by setting a section of a route that requires a mantle (term for using a heel and both arms to climb atop a ledge). Design criteria like this does not necessarily affect the difficulty of the route. 

\comment{
This claim appears unfounded but the logic is straight forward. The overall probability of success on a given route is given by the combined probability of each move in context on the wall. 

$$P(R) = \prod_{\forall r \in R} P(r)$$

Therefore, to decrease the probability of success on the route, $P(R)$, we may increase the number of constituent movements, $r$. So, given any movement or combination of movements we may create a route of arbitrary probability of success by increasing the length of the route by repeating movements. This is true whether the route is homogeneous or heterogeneous in the movements required.  

As a note, in a heterogeneous route, raising the grade (making the route more difficult) can be done by including one move that has a probability, $P(r \in R)$, which is very small. This is known as a one move route and is generally considered undesirable as the route is properly included in the appropriate grade but is not engaging overall. }

Since the style is not necessarily accounted for in the grade but is a reality of route setting, a computational route setter must be able to consider the desired style (movement content) of the route as well as the culminating grade. 

Climbers who have climbing experience that does not include any of a specific type of movement (non-uniform sampling of movement types) often find routes including these movements to be more difficult than others in the community who have more uniform exposure. Thus, grade disputes may arise from lack of exposure to a subset of movement types.

\comment{
In \cite{colton2011computational} the authors suggest the FACE formalism for generative/creative software. The creative act of the software is a 4 tuple that is made of $<F,A,C,E>$ (Framing, Aesthetic measure, Concept, Expression). The framing is expected to be some situating human readable text (this paper and our future work). The concept is the executable program (the end goal of our research). The expression is the artefact which is generated (the climbing route). And finally, the aesthetic measure takes the tuple $<C,E>$ and returns some positive value. However, in general this is reductive as the evaluation of some artefact may be multidimensional by nature. Indeed, for a climbing route generation system which aspires to human creativity and criticism, the evaluation necessarily includes the tuple $<g,e>$, the grade and the evaluation of the climbing experience as they are independent. Essentially, the human evaluation of the climbing route is some function that is a superposition of ortho-normal bases which includes but is not limited to the grade, climbing style, and visual style. 

From the application of the FACE model of creative software, it is clear that a software which is able to generate climbing routes in a manner which is consistent with human setters would represent a substantial work in the area of computational creativity and computational criticism. This establishes the intellectual merit of computational climbing route generation in the domain of computational creativity. 
}

\section{Literature Review}

We have numbered a set of desiderata for a computational agent that attempts to create or co-create rock climbing routes based on the discussion and analysis here.

 \begin{table}[h]
  \centering
  \caption{Desiderata for Climbing Route Generation Systems}
  \label{tab:climbinggendesiderata}
  \begin{tabular}{c p{22em}}
    \toprule
    \# & Desiderata\\
    \midrule
    - & generate a set of hold placements \\ \midrule 
    1 & predict (or be aware of) the sequence of movements humans will employ (path plan in vertical terrain) \\ \midrule 
    2 & predict the grade of the sequence of movements\\ \midrule
    3 & predict the reward from the sequence of movements (considering style) \\\midrule
    4 & adjust the hold placements such that the sequence of movements is more likely to conform to the desired grade and maximize reward  \\
    
    \bottomrule
  \end{tabular}
  
\vskip-0.5em
\end{table}

A literature review of all papers that have attempted computational route generation or grading has been conducted. Many of papers in the literature have a significant aspect in common: they ignore desiderata 1 which is fundamentally necessary for all future work in computational climbing route generation and grading. These papers are not discussed.

The most substantial work in the computational route generation literature, Strange Beta \cite{Phillips2011}, treats perturbing existing movement sequences (partial desiderata 1) as a creative process. Inspired by work in creating pleasing musical sequence variations\removeComment{\cite{dobson2007chaos}} and variations on choreographed dances\removeComment{\cite{bradley2010towards}}, the authors use chaos theory to produce creative variations on existing routes (a method which seems to have been little explored by the ICCC). This is the only paper in the literature that attempts to maximize the reward to the climber (desiderata 4). However, it does not attempt to reason about the grade or the reward (no attempt desiderata 2-3). 

Strange Beta (beta being a climber term for sequence of movements) requires route setters to encode routes in a domain specific language, called CRDL, to describe hand positions throughout the route. Then, the system uses chaotic Lorenz dynamics to produce variations of the route that were shown to be more rewarding to some climbers. This is an important computationally co-creative system that has not been discussed or referenced in the ICCC community. 

The authors of Strange Beta justify the use of hand movements to characterize a route by suggesting that foot holds are more incidental and may be added or removed to adjust the difficulty as necessary. They do note that this is not the case for ``exceptions". Considering the intent of this work was as a co-creative system that ignores grade, this simplification makes sense. 

In \cite{Stapel2020}, the authors engineer a set of descriptive heuristics based on additional, manually encoded information. They are able to capture information about the necessary movements (variation of desiderata 1) which they use, along with the order and distance between the holds, to predict the grade (attempt desiderata 2). The authors state they believe systems like theirs will be able to perform much better when a system is created that can predict the ``easiest order of holds" and to detect foot holds. This statement seems appropriate as the easiest order of holds and the appropriate foot holds would essentially constitute solving desiderata 1. This system is only applied to the moonboard climbing system which has a space of possible movements which is far smaller than is possible in climbing on traditional walls. 

\cite{Duh2021} uses hold positions and specifically calculates a difficulty score for each hand move by looking at the difficulty of a hold and the distance between holds. They essentially use a greedy search over a fully connected graph of the holds with the cost of each edge being based on distance and difficulty to predict the sequence of hand movements (attempt partial desiderata 1). The authors note that previous work has almost entirely ignored the movement of the climber when attempting to predict the grade. However, like previous work, they only attempt to predict the sequence of hand movements. This system is also only applied to the moonboard climbing system. 

The number of papers that have attempted route generation and grading is significant. However, the number of systems that have attempted to go beyond the hold locations is few. Of these none consider more than the movement of climbers' hands. 

Whether or not hand movements alone can be used to meaningfully characterize (grade and style) and generate climbing routes is an open question. However, it does not seem likely. Without considering a climber's feet, it is not possible to predict the position of the body, specifically the attitude and angle relative to the wall. \comment{Without knowing the complete initial condition, from the study of complex systems, it is clear that prediction is not possible.} To see that this is the case, consider the difference between climbing a ladder with feet on a lower rung versus having feet dangling in the air. The movement of the body (and hands) will be drastically different.

\section{Conclusion}

We have introduced computational route generation and grading to the computational creativity community and provided a basis for future work. We have proposed a grading algorithm which may prove to be more human consistent than the current best computational approach. We have identified the desiderata that a computational route setter should possess and shown that these are unsolved in the literature. Of significant note, no literature has attempted human consistent path planning in the climbing domain. We therefore consider desiderata 1 in Table \ref{tab:climbinggendesiderata} to be a critical goal for computational climbing route generation and grading. 

\removeComment{
Generating a set of hold placements is trivial assuming a computational system possesses a suitable domain specific language of expression. (1) can be viewed as a path planning problem in a subset of terrain that has not been considered in the path planning literature. This may be of less interest to the ICCC, but is of tremendous interest to the AI community at large. (2) - (4) are problems that we believe are interesting from a computationally creative perspective. (2) and (3) involve developing interesting computational critics that are capable of describing, interpreting, evaluating, and justifying \cite{roberts2020extending} a criticism of a climbing route. (4) employs the critics in a creative ecosystem to provide input to a search algorithm that seeks to find creative routes which meet certain goals.}

\bibliographystyle{iccc}
\bibliography{iccc}

\end{document}